**Scandium diboride: a semi-metallic, lattice, thermally matched substrate for vertical AlGaN power electronics**

MVS Chandrashekhar[1], Daniel Joel Harrison[1], Ahamed Raihan[1], Astrid D. Kengne[1], R. Shipra[1], Han Xie[2], Tasnia Jabin[1], Monte Hendrix[2], Ethan Scott[7], Roshan S. Annam[8], Sharad Mahatara[4], Evan N. Crites[1], Allana G. Iwanicki[2], Luke J. Meiler[2], Maxime A. Siegler[2], Renae Gannon[4], Stephen Spurgeon[4,5,6], Ashutosh Giri[8], Rajeswari Kolagani[9], Joshua A. Burrow[1], Stephan Lany[4], Patrick Hopkins[8], Tyrel M. McQueen[2], Michael Spencer[1], Satya Khushwaha[2,3]

[1]Microelectronics Education and Research Center, Morgan State University
[2]Chemistry and Physics, Johns Hopkins University,
[3]Physics, University of North Texas
[4]National Lab for the Rockies
[5]Metallurgical and Materials Engineering, Colorado School of Mines, Golden, CO
[6]Renewable and Sustainable Energy Institute (RASEI), U. of Colorado Boulder, Boulder, CO
[7]Mechanical and Aerospace Engineering, University of Virginia
[8]Mechanical Engineering, University of Maryland College Park
[9]Dept of Physics, Astronomy & Geosciences, Towson University

**Abstract:** We report the properties of hexagonal (space group P6/mmm) scandium diboride ($ScB_2$) single crystals grown by a laser diode floating zone method at growth rates of ~1mm/hr under B-rich conditions with (002) rocking curve widths $\Delta\omega$=38'' approaching the quality of commercial SiC/GaN substrates. Lattice expansion measurements reveal matching to $Al_{0.55}Ga_{0.45}N$ with a coefficient of thermal expansion ~5ppm/K at typical AlGaN growth temperatures, enabling thick AlGaN layers for ultra-wide bandgap (UWBG) power electronics >1kV. We measure semi-metallic room temperature resistivity ~$15\ \mu\Omega\ cm$, climbing to $\sim 93\mu\Omega\ cm$ at 773K with a $T^2$ dependence effectively eliminating substrate parasitic resistance, the limiting factor in exploiting the full potential of UWBG. The Debye temperature $\theta_{D,ScB2}$ from heat capacity and lattice expansion is ~850K well matched to $\theta_{D,AlGaN}$, but lower than the 1100K measured for Sc-rich growth conditions. We discuss Debye matching as a key substrate codesign criterion providing significant overlap in phonon modes for heat removal and thermal matching during AlGaN growth. The competitive thermal conductivity at room temperature 53W/mK is half that from first principles calculations, a discrepancy we attribute to the presence of Sc-vacancies generated by B-rich growth. while the resistivity is ~2x the theoretical value, indicating that both electrons and phonons play equal role in thermal transport. The smooth ~2.5nm rms roughness surface enables advanced heat removal modalities through engineered phonon bridges and phonon polaritons in $ScB_2$/AlGaN interfacial heterostructures, potentially allowing ~10-100x increase in power handling over state-of-the-art GaN/SiC.

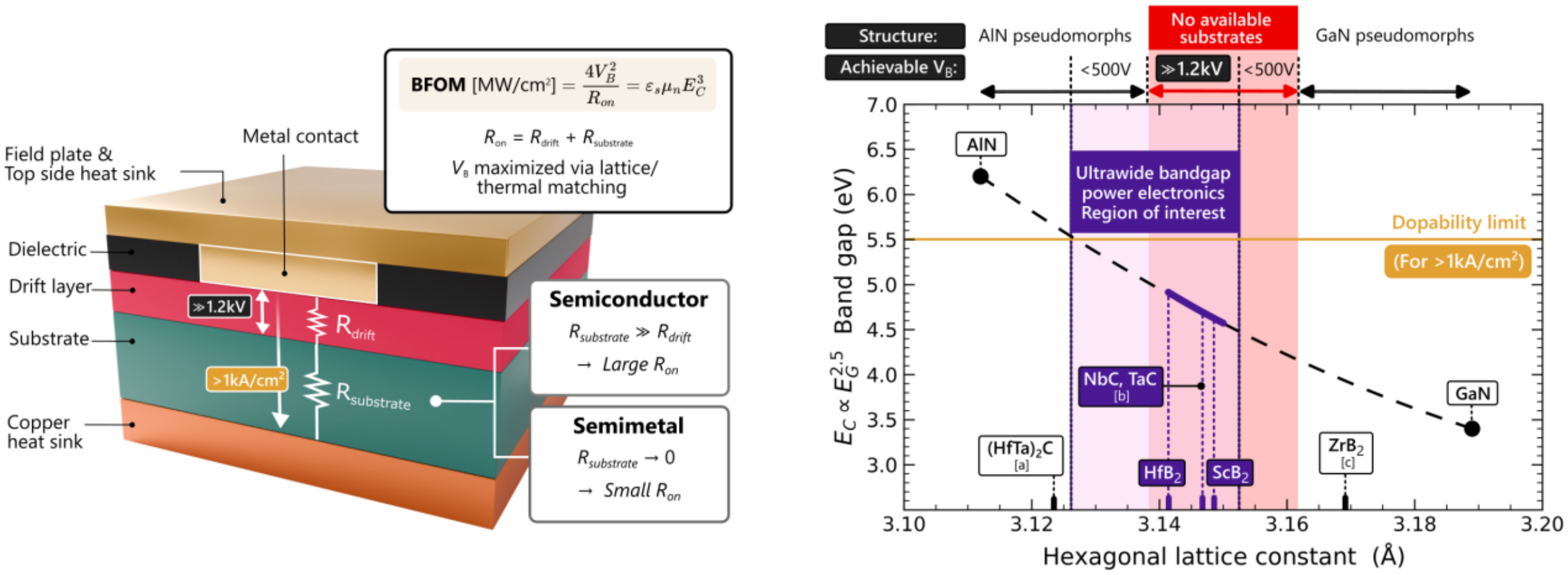

**Figure 1** Substrate availability and performance bottlenecks in ultrawide bandgap (UWBG >4eV) III-Nitride power electronics for continued miniaturization of power systems predicted by the Baliga Figure of Merit (BFOM). In addition to the Joule heating thermal load that must be managed, $R_{substrate}$ dominates the RC time constant. A longer RC time constant leads to slower pulse width modulation (PWM) switching frequencies, requiring larger passive elements (RLC), currently the key driver of the achievable size, weight and cost in large power systems. By migrating to semi-metal substrates with $\mu\Omega - cm$ vs m$\Omega - cm$ resistivities for conducting SiC/GaN/Si, the substrate bottleneck is eliminated. Furthermore, by choosing lattice and thermally matched semi metal substrates, strain in the epitaxial AlGaN is minimized, allowing very thick active layers to be grown. Overcoming the tyranny of the semiconductor substrate with semimetals enables $V_B$>10kV with current densities >1kA/cm2, exceeding the state of the art in GaN/SiC power devices. [a-c][1–3]

In today's energy landscape, high power/demand applications such as artificial intelligence (AI) data centers,[4] power grids,[5] advanced electric vehicles,[6] and consumer electronics, and their desire for high power/energy density current delivery systems has led to miniaturization of power electronic modules, leading to thermally limited performance limitations that has become endemic to high power density systems. Indeed, this thermal limitation applies to silicon complementary metal oxide semiconductor (CMOS) microelectronics for digital computation, including graphics processing units (GPU's) and microprocessors for AI, where Silicon's temperature must be managed very carefully as a tenuous balance near breakdown and thermal runaway to maximize performance. However, the deep Moore's law scaling of CMOS functionality in computation can only be achieved with Silicon presenting a necessary engineering constraint that is managed through added cost and complexity with multiple cores[7], advanced packaging strategies[8], and multimodal thermal management strategies such as vapor chambers, recirculating refrigerants and high surface area structures for passive air cooling.

However, in power electronics, there is no need for CMOS's deep level of integration for computation. Thus, to manage this thermal bottleneck for continued miniaturization, wide bandgap (WBG, energy bandgap $E_g > 3eV$ ) SiC and GaN electronics have begun to replace silicon power electronics devices in electric vehicles and consumer electronics owing to their higher temperature stability, and overall ability to handle high current densities >100Acm$^{-2}$ at elevated voltages beyond 1kV.[9,10] This replacement took many decades owing to material defects e.g. micropipes,[11] basal plane dislocations,[12] and surface trapping.[13] These defects led to reliability problems in the field.[14] WBG materials and doping technology matured, driven in no small part by the development of the blue LED that led to the 2014 Nobel prize in Physics, and now WBG power electronics find a foothold in power systems. WBG SiC/GaN are considered the gold standard owing to their higher electric field handling, temperature stability $> 350℃$, which lead to high current capacity and faster switching speeds all of which enable continued miniaturization in pulse width modulation (PWM) power delivery systems, the *de facto* standard[10].

These advances have led to ~3-5x increase in overall power density, which while significant, is still a far cry below the ~100-1000x expected from the Baliga figure of merit (BFOM) that codifies the power handling in an idealized vertical power Schottky diode (Figure 1) for a given material in terms of an application specified off-state breakdown voltage $V_B$ that determines the on-state resistance $R_{on}$ in terms of the semiconductor's dielectric constant $\epsilon_S$ electron mobility $\mu_n$, and critical breakdown electrical field $E_C \propto E_g^{2.5}$ a strong function of $E_g$ [10]

$$BFOM\left(\frac{W}{cm^2}\right) = \frac{V_B^2}{R_{on}} = \frac{V_B^2}{R_{drift,}+R_{substrate}} = \epsilon_S \mu_n E_C^3 \Rightarrow R_{substrate} \to 0 \qquad (1)$$

The final equality only holds if the parasitic substrate resistance $R_{substrate}$ is small compared to the active drift region resistance $R_{drift}$. For typical semiconductor substrates with resistivity $\sim 10m\Omega\, cm$, and a thickness of $500\mu m$, this presents a rather small $R_{substrate} \sim 1m\Omega\, cm^2$, although this begins to be a significant parasitic as specific on-resistances approach the $m\Omega\, cm^2$ level. In addition to limiting overall power handling, at a given current density $J_{on}$, this parasitic presents additional thermal load through Joule heating $J_{on}^2 R_{substrate}$ that must be removed to maximize the performance of the higher cost WBG module. Device designers have employed backside wafer thinning to reduce $R_{substrate}$[15] and advanced packaging strategies to handle higher temperatures[16] without which overall system power density cannot meaningfully improve.

Recently, ultrawide bandgap (UWBG) semiconductors such as AlN, AlGaN, $Ga_2O_3$, diamond, and cubic BN have emerged[17] as post WBG materials promising to improve power density by ~10-100x over SiC/GaN owing to even thinner active layers from their high $E_C > 10MV\ cm^{-1}$ enabling higher voltage to be held off in a thinner layer promising $R_{drift} < 10^{-4}\Omega\, cm^2$. However, at these low resistances, the parasitic $R_{substrate}$ becomes a killer bottleneck particularly for near term voltage nodes $V_B = 1.2 - 3.6kV$ providing no meaningful improvement over WBG. Moreover, doping for UWBG is a challenge due to the deep donor levels and complex defect chemistries that emerge such as with AlN (DX centers and oxygen impurities)[18,19], limiting the near-term prospects to semiconductors with $E_g < 5.5eV$ (Figure 1), as dopability is necessary to generate free carriers to minimize $R_{drift}$. With doping constraints there are no lattice-matched traditional semiconductor substrates, without which defect-free thicker layers cannot be realized for $V_B > 1.2kV$. As a typical example, for $Al_xGa_{1-x}N$ (x~0.5) with $E_g \sim 5eV$,[20] a 1kV device requires ~$2\mu m$ thick active layers doped with donors at $N_D \sim 10^{17} cm^{-3}$ via the relationship $N_D = 2\epsilon_S E_C^2 / 4qV_B$.[10,20] Growth on the best matched native bulk AlN substrate is still mismatched enough to limit active layer thickness to $\sim 0.5\mu m$, providing no advantage over Si, let alone WBG SiC/GaN. The lack of appropriately lattice and coefficient of thermal expansion (CTE) matched[2,3,21] low $R_{substrate}$ substrates for UWBG constitutes a key limitation in further scaling of power electronic devices and systems.

This tyranny of the substrate with traditional semiconductor bulk wafers can be broken by using semimetals such as TaC,[2,20] $(Hf\text{-}Ta)_2C$,[1] and $ZrB_2$, all of which have been measured to be lattice matched to $Al_xGa_{1-x}N$ with TaC and $(Hf\text{-}Ta)_2C$ for $x \sim 0.5 - 0.7$, while $ZrB_2$ is a closer match to x~0.2 (Figure 1). Molecular beam growth of unstrained epitaxial $Al_xGa_{1-x}N$ (x~0.6) was demonstrated on semimetallic TaC (bulk resistivity $37\mu\Omega\, cm$)[2] virtual substrates,[2] although the development of bulk TaC[1] is required for true vertical devices. Significantly more work was done with GaN films on bulk $ZrB_2$ grown by metalorganic chemical vapor deposition (MOCVD) demonstrated by one of the 2014 Nobel groups.[3,21] However, the emergence of bulk GaN substrates partially obviated the need for exotic substrates at that time.

Transition metal diborides $MB_2$ have generated great interest since the discovery of superconductivity in $MgB_2$.[22,23] Since then, many studies on synthesis of these layered hexagonal ultrahard $MB_2$ ceramics and composites[24] for mechanical applications have shown metallic resistivities in the $\sim 10\mu\Omega\, cm$ range for these refractory semimetals with melting points exceeding 3000℃ for $HfB_2$, a tantalizing prospect for power device substrates particularly considering the success of $ZrB_2$ for GaN. Topological transport properties have been suggested in $HfB_2$,[25,26] while de Haas van Alphen oscillations were observed in $HfB_2$, $ZrB_2$ and $ScB_2$. Strain tunable superconductivity has been predicted for $ScB_2$[26]. The potential topological properties in $MB_2$ combined with their suitability as UWBG substrates could find advanced power applications such as superconducting interconnects in data centers[27,28] and magnetic levitation trains leveraging He-4 cooling.[29]

Bulk $ScB_2$ with a melting point of ~2200℃, was grown under Sc-rich conditions using floating zone growth by Levchenko et. al,[30] leading to follow-on transport studies[31] as well as theory predictions[32] of the electronic properties, all of which culminated in a detailed review of the thermoelastic properties of various borides.[33] In those studies, $ScB_2$ emerged as the boride with the highest degree of anisotropy, with a measured 2x difference in effective mass between the cardinal (001) and (100) directions, while ~30% anisotropy was predicted for electrical resistivity. Based on their lattice constant measurements, we predict a lattice match to $Al_xGa_{1-x}N$ (x~0.55) in Figure 1.

In this paper, we demonstrate the suitability of high-quality Scandium Diboride ($ScB_2$) single crystals grown under B-rich conditions by a laser diode floating zone technique as ideal scalable substrates for $Al_xGa_{1-x}N$ (x~0.55) vertical power devices to unleash the full promise of this UWBG semiconductor alloy for power electronics and discuss the key issues that drive the choice of substrate in electrothermally codesigned power devices.[34,35]

**Experiment and Methods**

The $ScB_2$ crystals were grown in a tilted laser diode floating zone furnace (LDFZ) using a $5 \times 200W$ 974nm laser system.[36,37] The growth geometry used a 6mm diameter seed rod of 99.9% pure commercially sintered $ScB_2$ and a similar feed rod, with a flux pellet in between to create a molten zone. Flux pellets were generated by mixing ultra-pure Boron powder with powdered $ScB_2$ rods and pressed at 5000psi in a Carver hydraulic press. In this paper, the most reproducible highest quality crystals were obtained by growing on the B-rich side of the phase diagram in contrast with the only other report in the literature[30] using a target flux of $B_{0.82}Sc_{0.18}$ at atmosphere pressure under Ar flow of 0.5slm at a laser power 70-80%. Typical growth rates were between 1 – 1.5 mm/hr with rotation ~10RPM. Further details of the growth will be reported elsewhere[38].

The grown crystals were coarsely oriented using Laue panchromatic x-ray diffraction in a Multiwire Back-Reflection Laue Camera System. The roughly oriented crystals were then bonded to a specialized goniometer for crystal slicing using Crystalbond wax, and then re-oriented as above. These finely oriented crystals were then cut using a diamond wire saw and polished to a mirror finish with diamond grit from $30\mu m$ down to $0.1\mu m$.

Single crystal x-ray diffraction (SXRD) was performed on a piece of $ScB_2$ single crystal approximately $\sim 20\mu m$ in size mechanically cleaved from the Boule using a Rigaku XtaLAB Synergy-R diffractometer equipped with a rotating-anode X-ray source, using Cu Kα radiation (λ = 1.54178 Å) and a HyPix-6000HE detector. Cell refinement and data reduction were performed with the program CrysAlisPro[39]. The temperature of the data collection was controlled using the system Cryostream 1000 from Oxford Cryosystems. Lattice constants were thus obtained as a function of temperature for heating and cooling cycles from 103-293K. Complementary high resolution X-ray diffraction (HRXRD) rocking curve (RC) measurements were performed on large polished single crystals using Rigaku SmartLab diffractometer. RCs were collected using the ω- scans between relative positions of -1 to +1 to the ω (~ 26.09(1)°) position of the (002) plane of $ScB_2$ and corrected for the measured equipment broadening of 24''.

For van der Pauw resistivity measurements, approximately $150\mu m$ thick slivers ~$1mm \times 1mm$ were mechanically cleaved and mounted in an Instec HCP621G-PMH environmental Hall measurement chamber with a sapphire insulator for electrical isolation from the chuck, with four gold coated system probes directly contacting the $ScB_2$ per the NIST standard[40]. Current levels from 0.1A to 1A, supplied by a Keithley 2450 source meter, were used to ensure linearity of the measurement and to preclude self-heating effects in this highly conducting sample. Heat capacity measurements from T = 2-300 K were taken with a Quantum Design Physical Properties Measurement System (PPMS) using the semi-adiabatic pulse method with a 1% temperature rise using a 1.4 mg sample. Prior to the measurement of actual sample, an addendum was measured using a small amount of Apiezon N grease.

The TDTR measurements[41] were performed in a similar manner as described in Ref. [42], in a two-tint configuration using a pulsed Ti:Sapphire laser with a pulse width of 150 fs, a repetition rate of 80 MHz, a central wavelength of 808 nm, and a full width at half maximum (FWHM) of 14 nm. The pump beam was modulated at 8.4 MHz, and the probe pulses were time-delayed relative to the pulses of the pump with a linear delay stage, providing up to 5.5 ns of delay. The focused $1/e^2$ diameters of the probe and pump beam were 12 and 19 μm, respectively. To extract thermal properties, we analyze the ratio of the in-phase and out-of-phase lock-in signals using a thermal model where parameters of interest are obtained via least-squares optimization. We model the sample as a two-layer stack: (1) an 80 nm thick Al transducer upon (2) a semi-infinite $ScB_2$ substrate. The thickness and thermal conductivity of the Al transducer are characterized from a witness sample included in the same deposition as the $ScB_2$, and found to be $80 \pm 3\ nm$ and $150 \pm 5\ Wm^{-1}K^{-1}$, respectively.

First principles calculations of the electronic density of states were performed with the projector augmented wave (PAW) method implemented in the VASP code[43] using the Strongly Constrained and Appropriately Normed (SCAN) meta-GGA functional[44] (SI-C). First principles calculations of the electrical resistivity were carried out using density-functional theory (DFT) and density-functional perturbation theory (DFPT) methods, as implemented in the Quantum Espresso suite, while those of thermal conductivity were obtained via the linearized phonon Boltzmann transport equation solved iteratively using ShengBTE (SI-G).

## Results and Discussion

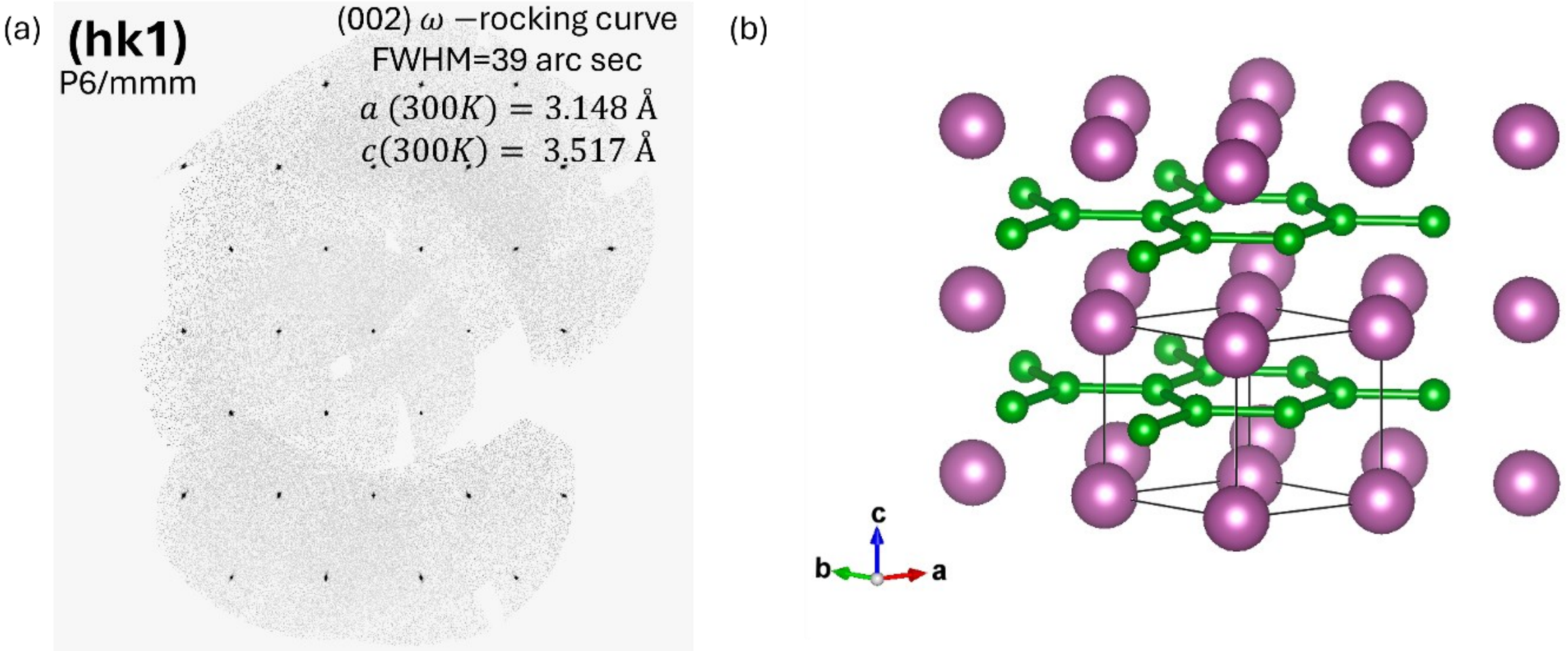

**Figure 2 (a)** Room temperature single crystal x-ray diffraction (SXRD) pattern for (hk1) reflections from which the crystal structure of $ScB_2$ was refined. We obtain a room temperature lattice constant of a=3.148Å and c=3.517Å, an excellent match for (001) $Al_xGa_{1-x}N$ x~0.55 **(b)** measured unit cell of $ScB_2$ with Sc as purple spheres and B as green spheres.

Figure 2a shows a reciprocal space slice measured for the (hk1) family of reflections by SCXRD where the hexagonal symmetry is clearly observed in SCXRD as well as in the 6 nodes in the HRXRD $\phi$-scan of the asymmetric (101) reflection. A detailed refinement of the structure revealed a space group symmetry P6/mmm with room temperature lattice constants $a = 3.148\text{Å}$ and $c = 3.517\text{Å}$, in excellent agreement with Levchenko et al.[30] Typical (002) HRXRD $\omega$ rocking curves give a FWHM $\Delta\omega \sim 38''$ (SI), approaching the quality of commercial SiC/GaN substrates[45,46]. This quality is supported by the sharp spots in the reciprocal lattice (Figure 1a). In the Debye model of lattice expansion, the temperature T dependence of the lattice constant is given by:[47]

$$V = V_0 + I_v T F(\theta_D/T) \qquad \text{(2a)}$$

$$F\left(\frac{\theta_D}{T}\right) = 3\left(\frac{T}{\theta_D}\right)^3 \int_0^{\frac{\theta_D}{T}} \frac{x^3}{e^x - 1} dx \qquad \text{(2b)}$$

where $\theta_D$ is the Debye temperature that controls lattice expansion. A fit to this model using equation (2) (Figure 3b) gives $\theta_D = 850K$, lower than the 1100K obtained from ultrasound measurements on crystals grown under Sc-rich conditions by Levchenko et. al[30] and in this work where Sc-rich growth did not yield large crystals suitable for substrates (see SI). Based on the full band theoretical calculations of the phonon density of states,[32] where the highest energy phonons are ~100meV corresponding to an effective phonon temperature ~1150K, we would expect $\theta_D$ to be in that range. We will see later that this reduced $\theta_D = 850K$ for $ScB_2$ with B-rich growth conditions is consistent with that extracted from heat capacity (Figure 4), suggesting that Sc-vacancies are responsible for softening the structure suppressing $\theta_D$. These results suggest that Sc-vacancies are more likely to form under B-rich conditions than B-vacancies under Sc-rich conditions. This asymmetry could arise from the higher energy cost associated with creating B-vacancies compared to Sc-vacancies due to the smaller B-atom size. Such asymmetry in vacancy formation energy is seen in the similar structure diboride P6/mmm $MgB_2$ ~4eV for B-vacancies and ~2eV for Mg-vacancies,[48] although the situation appears to be reversed for the apparently similar P6/mmm $ZrB_2$,[49] indicating that further investigation is required.

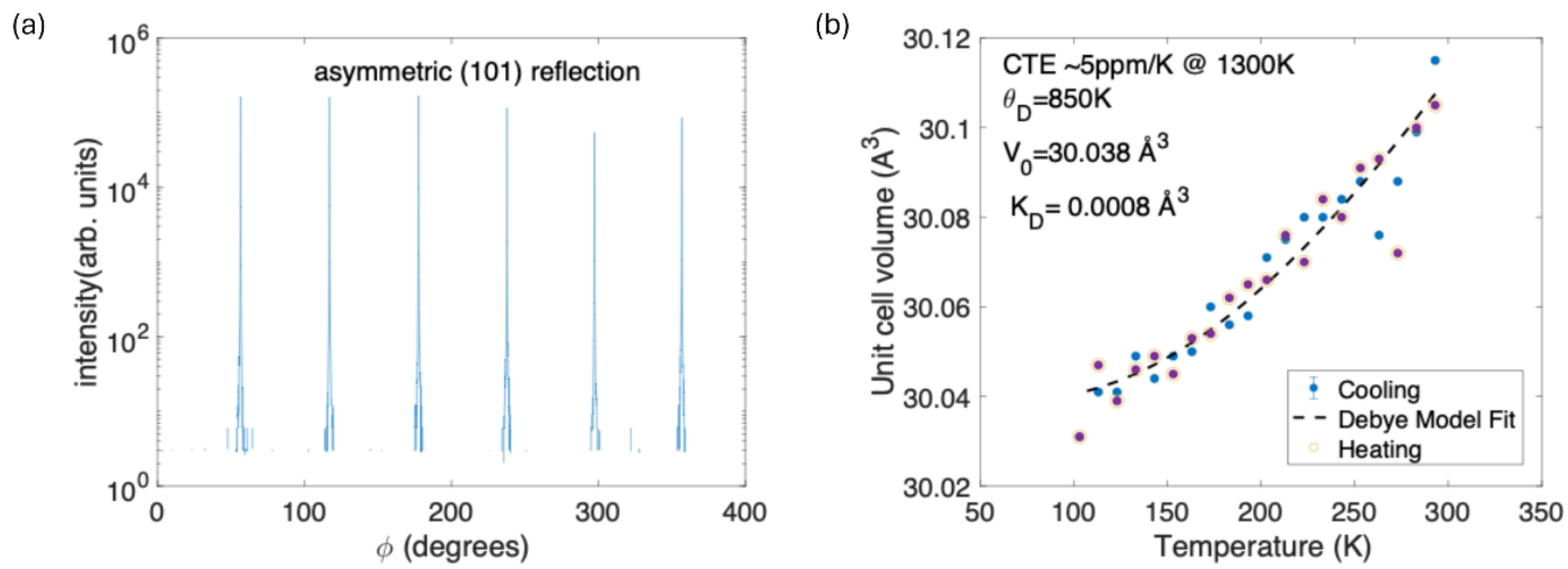

**Figure 3 (a)** Asymmetric (101) reflection ϕ-scan from HRXRD with the 6-node hexagonal symmetry of the c-plane clearly visible (see SI for additional reflections). **(b)** Temperature dependence of unit cell volume indicating a Debye temperature $\Theta_D$ ~850K, corresponding to an isotropic coefficient of thermal expansion (CTE) ~5ppm/K, a good match for $Al_xGa_{1-x}N$ x~0.55,[50,51] a necessary condition for the growth of thick AlGaN epitaxial layers for high voltage blocking power electronics.

The measured lattice constant $a$ =3.148Å for $ScB_2$ matches that $Al_xGa_{1-x}N$ x~0.55 [52] (Figure 1) in line with Levchenko et. al.[30] From this temperature dependence, we can extrapolate the coefficient of thermal expansion (CTE) at typical $Al_xGa_{1-x}N$ growth temperatures[2] 1000℃ ~5ppm/K assuming hydrostatic expansion of this bulk crystal, which is also an excellent match for $Al_xGa_{1-x}N$ (x~0.55)[52]. The thermal match is not surprising given the relatively close match of $\theta_D \sim 850K$ between $ScB_2$ and $Al_xGa_{1-x}N$ (interpolated between GaN and AlN) as compared to the mismatch between TaC and $Al_xGa_{1-x}N$ (x~0.55) (Figure 6a), given that $\theta_D$ drives lattice expansion in the Debye picture. This points to the desirability of matching $\theta_D$, a point which will be underscored further below in the discussion of heat capacity, and in the broader discussion below on substrate matching to the active drift layer.

Electrical resistivity: The electrical resistivity $\rho$ in metals increases with temperature due to increased scattering from intrinsic phonons and electrons, and extrinsic impurities and defects. For phonon scattering, $\rho$ scales with $\theta_D$ via the Bloch-Gruneisen (B-G) relationship,

$\rho = \rho_0 + C \left(\frac{T}{\theta_D}\right)^n \int_0^{\theta_D/T} \frac{x^n}{(e^x+1)(1-e^{-x})} dx$, where we take n=5 for 3-dimensional metals[53], and $\rho_0$ is the residual resistivity. Figure 4a shows the temperature dependence of resistivity. The low temperature resistivity $\rho_0$ approaches $13\mu\Omega\ cm$, a residual resistivity we attribute to trace impurities in the crystal, similar to that observed in metals.[54] This residual value is~4x higher than that predicted by Sichkar using a variational method,[32] and ~20x higher than our first principles calculations. The difference in predicted resistivity arises from a different electron phonon coupling mass enhancement parameter $\lambda_{e-ph}$~$0.28$ in our calculations as compared to 0.48 in Sichkar's work.[32] At room temperature, the measured resistivity, still near the residual value has a smaller discrepancy with theory, underscoring the high quality of these crystals. Further reduction in impurities may allow the lower resistivities and quantum coherence[25,31] of these crystals to be accessed. In the B-G model, $\rho \propto T^5$ is expected at intermediate temperatures below $\theta_D$. In our measurements however, at temperatures above 400K $\rho \propto T^2$ is observed in contrast with Sc-rich grown samples (SI F), suggesting that electron-electron scattering dominates[55] at elevated temperatures. The origin of $\propto T^2$ behavior may be related to the B-rich growth conditions leading to Sc-vacancies and related defects to be investigated in future work. Nevertheless, even at 500℃, the resistivity is$< 100\ \mu\Omega\ cm$, significant enough to erase $R_{substrate}$ in equation 1 in high power density environments.

From spectroscopic ellipsometry on (100) $ScB_2$, the plasma frequency $\omega_p$ was measured to be $\hbar\omega_p = 1.4eV$ (SI) in the infrared as compared to metals that have deep ultraviolet $\omega_p$. Within the Drude picture, we estimate a room temperature low frequency/DC $\omega_p\tau = 1/\rho\epsilon_0\omega_p = 270$ indicating very low losses, making $ScB_2$ suitable for plasmonic applications,[56] particularly with lattice matched III-Nitrides. These low losses also provide for thermal management using phonon polaritons[57] for heat removal away from device hot spots given the overlap in the $ScB_2$ phonon spectrum[58] with LO/TO $Al_xGa1_{1-x}N$ (x~0.55) phonons $A_1(LO)$~100meV and $E_2$~75meV.[58] This potential for plasmonic thermal management using lattice matched $ScB_2$ substrates provides an additional advantage for III-Nitride power design co-design.[34,35]

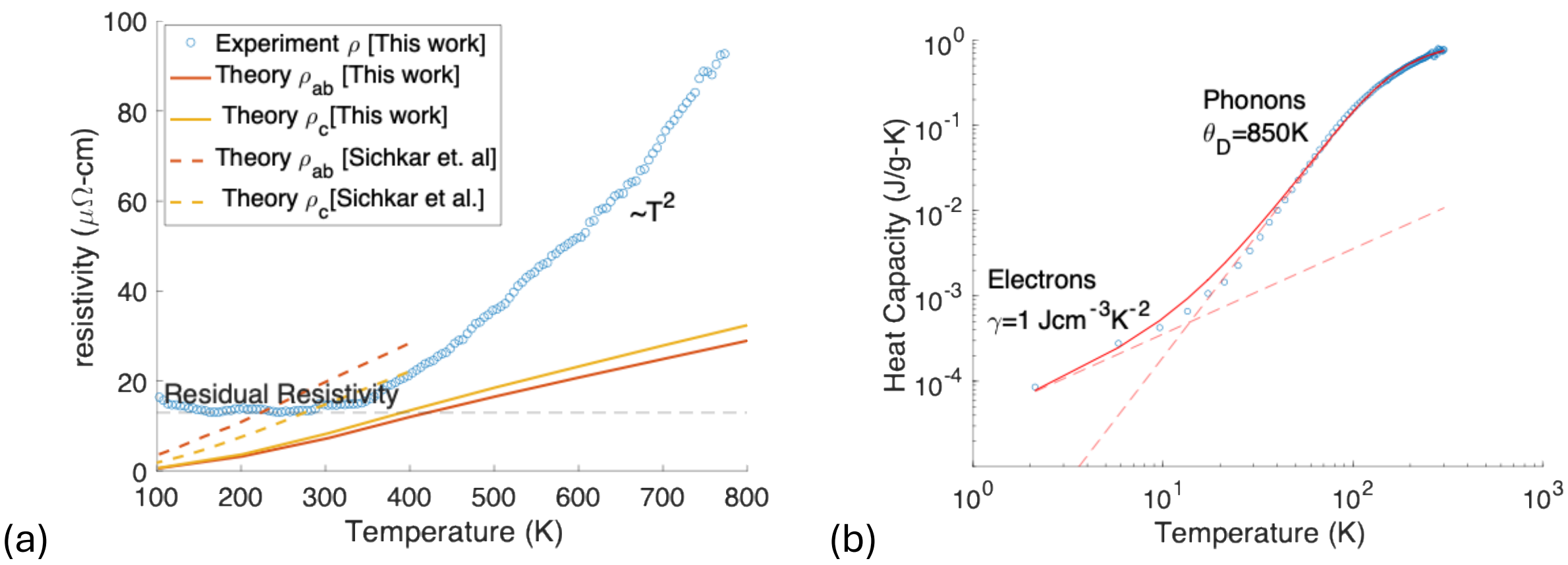

**Figure 4** Temperature dependence **(a)** of bulk resistivity showing $\sim 13\ \mu\Omega\, cm$ residual resistivity and a $T^2$dependence indicating that electron-electron scattering dominates[55], giving a higher resistivity than with electron-phonon scattering from theory. **(b)** heat capacity showing a Debye temperature $\Theta_D$~850K in agreement with lattice expansion measurements in Figure 2.

Heat capacity: The heat capacity at constant volume $C_v$ is a sum of electronic $C_{el}$ and phonon $C_{ph}$components[59]

$$C_v = C_{el} + C_{ph} = \gamma T + 3k\left(\frac{T}{\theta_D}\right)^3 \int_0^{\frac{\theta_D}{T}} \frac{x^4 e^x}{(e^x-1)^2} dx \qquad (3)$$

Where $\gamma = \pi^2/3\ k_B^2 D(E_F)$ is the Sommerfeld coefficient for electronic heat capacity[60], and $D(E_F)$ is the density of states at the Fermi level $E_F$. Despite the lattice expansion described above, constant volume is a reasonable approximation commonly used in solids giving a small error <2%.[59] Figure 4b shows $C_v$ as a function of T, where the electronic and phononic regimes are clearly delineated, with $C_{el}$ dominating below ~10K with a linear dependence and a measured $\gamma = 1.3 \times 10^{-4} JK^{-1}cm^{-3}$. From the theoretical calculations of the band structure (see SI Figure C), $D(E_F) = 28.3 eV^{-1}nm^{-3}$, giving an estimated value $10^{-4} JK^{-1}cm^{-3}$ in excellent agreement with our measurements. Above ~10K, $C_{ph}$ dominates, following a $\sim T^3$dependence as expected in the Debye model of heat capacity, and a fit to equation (3) reveals a characteristic $\theta_D = 850K$, consistent with lattice expansion, but lower than for samples grown under Sc-rich conditions,[30] similar to the trend for lattice expansion described above (see SI), which we tentatively attribute to lattice softening from Sc-vacancy formation under B-rich conditions.

Thermal conductivity: Figure 5a shows the TDTR signal for off-oriented (121) surface identified by electron backscatter diffraction (EBSD)[61] (SI Figure E) as a single phase after finishing with an ion beam of an ~2.5nm rms roughness optically polished surface (SI Figure D). This orientation was used to determine anisotropy in thermal conductivity $k_{th}$ through time domain thermoreflectance (TDTR) in comparison with the basal a-plane with (200) $\omega$ rocking curve FWHM $\Delta\omega = 132''$ (see SI). For this basal a-plane, heat generated from the laser being absorbed by the Al transducer is transported perpendicular to this plane along the (001) c-plane. We assume a value of $2.42\ MJm^{-3}K^{-1}$ for the volumetric heat capacity of the Al, and apply a measured value of $2.6 \pm 0.2\ MJm^{-3}K^{-1}$for the $ScB_2$ substrate (Figure 4b). The Al/$ScB_2$ interface and thermal conductivity of the $ScB_2$ are then treated as fitted parameters. The assumed parameter uncertainty is obtained through a Monte Carlo simulation which accounts for the uncertainty in the non-fitted parameters. We note that the interfacial thermal conductance G at the Al/$ScB_2$ interface $40 - 60MW\ m^{-2}K^{-1}$ is explicitly decoupled in this TDTR analysis (Figure 5a-d), so that the true bulk $k_{th}$is extracted. This approach was repeated for several crystals yielding bulk $k_{th}$ thermal conductivities of $53 \pm 5\ Wm^{-1}K^{-1}$ for (121) oriented samples and $53 \pm 6\ Wm^{-1}K^{-1}$for a (100) oriented sample (Figure 5b) showing that the isotropy predicted from our first principles calculations (Figure 5e, SI Section G) is indeed observed, in contradiction to the anisotropy suggested in previous studies.[32,33]

However, the total measured $k_{th}$ is ~50% of the theoretical value >100W/mK at room temperature, another clear indicator of lattice softening possibly from Sc-vacancies reflected in the suppressed $\theta_D{\sim}850K$ despite the high quality of these single crystals (Figure 2). Figure 6a compares $k_{th}$ as a function of $\theta_D$ for semiconductors and semimetals with phonon dominated thermal conduction with those for metals. $ScB_2$ shows overall $k_{th}$ comparable to that expected for phonon dominated thermal conduction as described by the Leibfried-Schlomann expression $k_{th} \propto \theta_D^3$.[62] However, this belies the fact that both electronic and phononic contributions are impacted. The room temperature electronic resistivity is ~2x the theoretical value (Figure 4a), indicating that the contribution to $k_{th}(electronic)$ is reduced by 1/2 to ~30W/mK. Similarly, from the Leibfried-Schlomann expression, the phononic contribution $k_{th}(phononic)$ is reduced due to the lattice softening by $(1100K/850K)^3$ to ~25W/mK, giving a total $k_{th}{\sim}55W/mK$ in excellent agreement with our measured values (Figure 5b). Thus, it can be inferred that there are equal contributions to thermal conductivity from electrons and phonons in our $ScB_2$ crystals. We expect that by further optimization of growth conditions to manage Sc-vacancies to increase $\theta_D$ to 1100K as expected from theory and Sc-rich growth conditions (SI and Levchenko et at.[30]), that a doubling to >100W/mK is achievable.

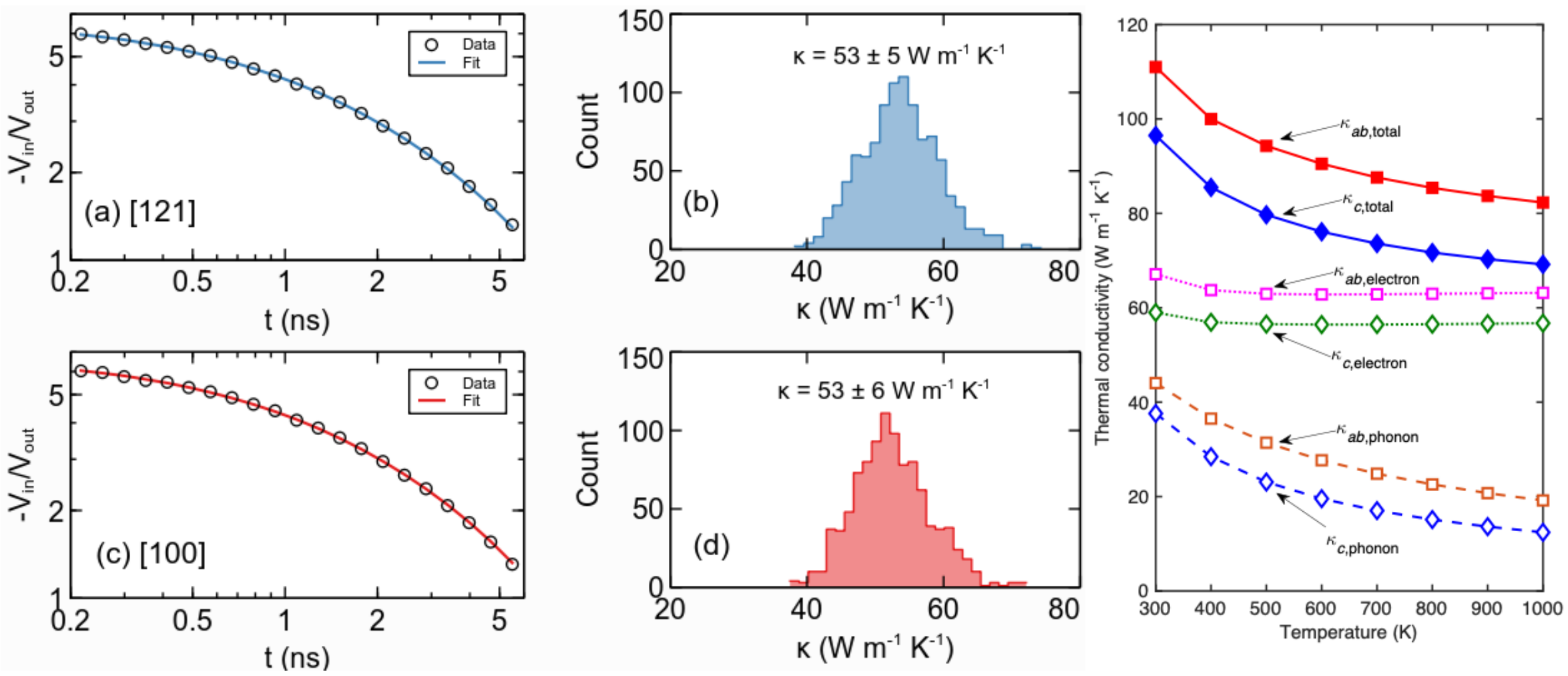

**Figure 5** (a) shows an example fit of time domain thermoreflectance TDTR ratio data for the [121] oriented crystal. Monte Carlo simulations are used to establish uncertainty in the fitted thermal conductivity as shown in (b). An example of the same analysis is shown of the [100] crystal in (c) and (d). (e) First principles calculations from the structure measured in Figure 2, showing weak anisotropy, and comparable contributions from electrons and phonons, placing the semimetal $ScB_2$ squarely between semiconductors (phonon dominated), and metals (electron dominated via the Wiedemann-Franz relation). The measured $k_{th}$ is ~50% of the theoretical value attributed to a suppressed Debye temperature $\theta_D$ reducing the phononic contributions and higher resistivity (Figure 4) reducing the electronic contributions.

Matching of $\theta_D$ between substrate and epitaxial layer can provide reduced thermal interface resistances based on phonon spectrum similarities[63–66], particularly given the smooth surfaces with rms roughness ~2.5nm (SI Figure D). This $\theta_D$ matching also provides for CTE matching as discussed above for managing strain during epitaxial growth, as well as during heating/cooling during subsequent fabrication of power devices. This matching condition, while not explicitly called out, is clear in a few key examples. AlN and sapphire have $\theta_D \sim 1100K$, which likely explains their common mating for synthesis of templates in III-Nitride microelectronics, and their surprising robustness across many growth conditions for various LED and power devices, as well as operation in the field. Another closely related example is between $ZrB_2$ and GaN[21], where the thermal matching during growth was called out as an advantage by measurement without explicit mention of $\theta_D$. On the other hand, the recent demonstration of AlGaN ($\theta_D \sim 850K$)[67] growth on lattice matched TaC ($\theta_D \sim 450K$)[68] templates,[2] which while promising, could cause reliability issues for power devices from thermal cycling from post-growth processing and field operation owing to the significant $\theta_D$ mismatch, in addition to the likely poor overlap in phonon modes leading to poor heat removal from the AlGaN/TaC interface. In other words, in addition to lattice matching at room temperature, the choice of substrate for growth of thick drift layers in high voltage power electronics requires consideration of $\theta_D$ matching, a previously unappreciated constraint that should become a key codesign criterion.[34,35]

In addition to the matching conditions above, the ideal substrate should have very low electrical resistivity to increase current handling and minimize parasitic substrate Joule heating, and high thermal conductivity to dissipate any heat that is generated in the device. Figure 6b compares the thermal conductivity of these same substrate materials, showing that the use of semi-metallic substrates such as $ScB_2$ reduces the substrate parasitic resistance by *3 orders of magnitude*. Even with the small penalty from suppressed $\theta_D$ and increased resistivity (Figure 4), this is still transformative to overall power handling particularly for UWBG device with increasingly lower $R_{drift}$ that can finally translate to overall lower $R_{on}$ (Figure 1).

Finally, we note that manufacturing practicality impacts scaling of substrate technology. For example, thermally limited UWBG substrates such as $Ga_2O_3$ have gained favor simply due to the fact that they can be grown from the melt, despite having other sub-par properties such as resistivity as compared to GaN/SiC. This tradeoff is accepted as melt growth brings scaling strategies from the silicon roadmap, a crucial factor in large scale deployment. $MB_2$, particularly $ScB_2$ are all melt grown, and therefore can adopt the same scaling roadmap, which along with the matching criteria above bring the full promise of UWBG semiconductors to power electronics in compact, efficient systems.

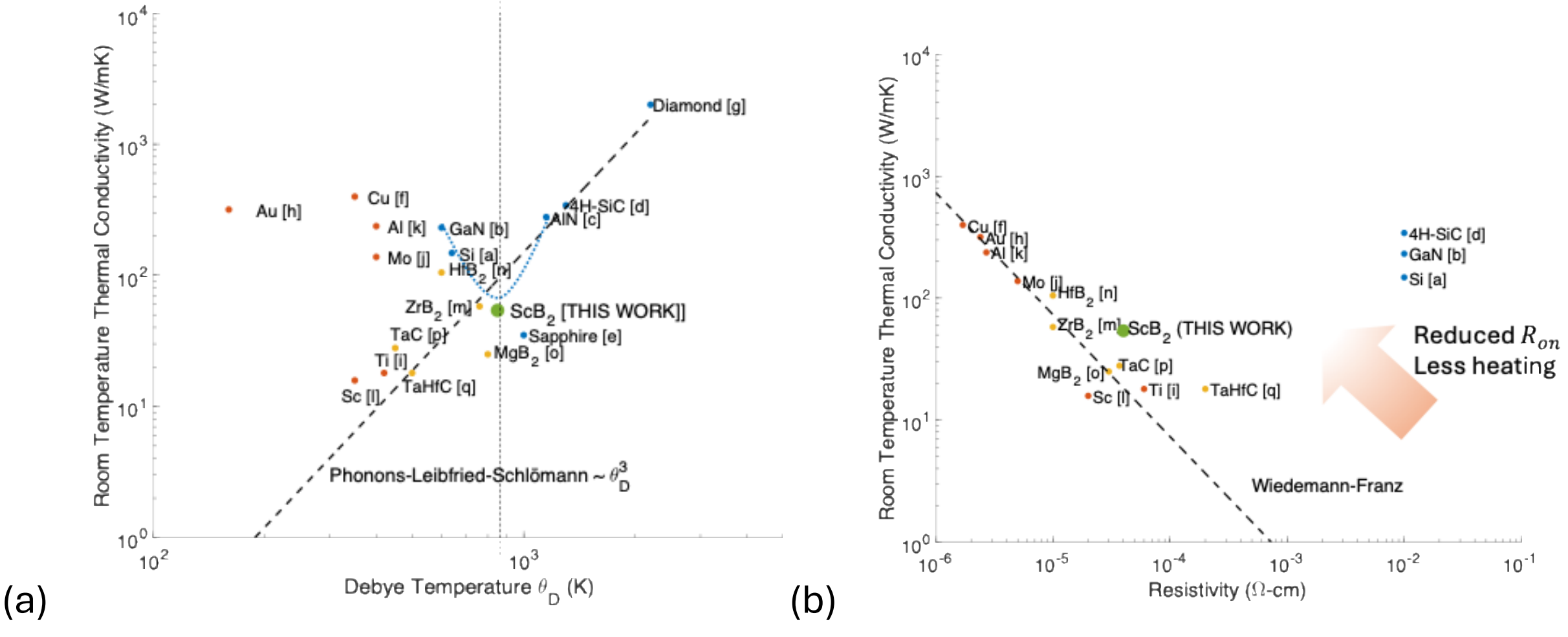

**Figure 6** Comparison of **(a)** thermal conductivity $k_{th}$ vs. Debye temperature $\theta_D$ with electrical resistivity and phonon mean free paths indicated and **(b)** $k_{th}$ vs. electrical resistivity $\rho$, of various metals, semiconductors, and semimetals, showing the unique capabilities that $ScB_2$ brings as a substrate for power electronics. [a-q][69–82,51,1]

In summary, we have demonstrated the suitability of high-quality bulk $ScB_2$ single crystals grown by laser diode floating zone method under B-rich conditions as a lattice and thermally matched substrate for growth of epitaxial $Al_xGa_{1x}N$ (x~0.55) vertical power electronic device structures. This matching allows thick layers to be grown for high voltage devices >1.2kV currently inaccessible for practical dopable $Al_xGa_{1x}N$ alloys, while the $\rho(300K) = 15\mu\Omega\ cm$ metallic resistivity eliminates the parasitic substrate resistance $R_{substrate}$ and consequent Joule heating thermal bottleneck with traditional semiconductor substrates. This low resistivity implies low loss in the infrared suitable for phonon polaritonic heat transfer at the AlGaN/$ScB_2$ interface. The measured isotropic thermal conductivity ~53W/mK is competitive considering the elimination of the substrate thermal load and is among the highest of any semimetal, although we speculate it is reduced from the theoretical value of 100W/mK possibly due to the presence of Sc-vacancies softening the lattice. A comparison of $ScB_2$ in the substrate landscape reveals that matching of substrate/active drift layer $\theta_D$~850K for $ScB_2$/$Al_xGa_{1-x}N$, is a key criterion that provides CTE matching to minimize strain related defects during epitaxial growth, and phonon spectrum overlap for efficient interfacial thermal transport, all essential for high BFOM codesigned power electronic devices.

**Acknowledgements**

This work was primarily supported as a part of APEX: A Center for Power Electronics Materials and Manufacturing Exploration, an Energy Frontier Research Center funded by the U.S. Department of Energy, Office of Science (growth, characterization, calculations analysis). The first principles calculations at University of Maryland (UMD) were supported by the Office of Naval Research, Grant No. N000142612113. UMD authors acknowledge the University of Maryland supercomputing resources, including the Zaratan high-performance computing cluster (https://hpcc.umd.edu), made available for conducting the research reported in this paper. Facilities management at Johns Hopkins University was supported by the National Science Foundation (Platform for the Accelerated Realization, Analysis, and Discovery of Interface Materials (PARADIM)) under Cooperative Agreement No. DMR-2039380. The Morgan State University authors acknowledge Morgan Center for Education and Research in Microelectronics (MERC), supported by the State of Maryland, for equipment, space, and technical support staff. This work was authored in part by the National Laboratory for the Rockies (NLR) for the U.S. Department of Energy (DOE), operated under Contract No. DE-AC36-08GO28308. The views expressed in the article do not necessarily represent the views of the DOE or the U.S. Government.

Supplementary Information

## A. Ellipsometry for plasma frequency measurement

**Spectroscopic ellipsometry** measurements were performed on the (100) $ScB_2$ single crystal at room temperature 300 K using an RC2 spectroscopic ellipsometer (J.A. Woollam Co., Inc.) at an angle of incidence of 60°. The ellipsometric parameters $\psi$ and $\Delta$ were measured over a photon energy range of approximately 0.5–5.9 eV. The experimental spectra were analyzed using CompleteEASE software and fitted using an optical model consisting of a Drude free carrier contribution and two Gaussian oscillators. A surface roughness layer was included in the model. The resulting model reproduces the measured $\psi$ and $\Delta$ spectra and was used to obtain the complex dielectric function, $\widetilde{\varepsilon} = \varepsilon_1 + i\varepsilon_2$. The plasma frequency $\omega_p$ was identified to be $\hbar\omega_p = 1.4eV$, corresponding to $\omega_p \approx 2.21 \times 10^{15}\ rad\ s^{-1}$ and a wavelength of approximately 854 nm.

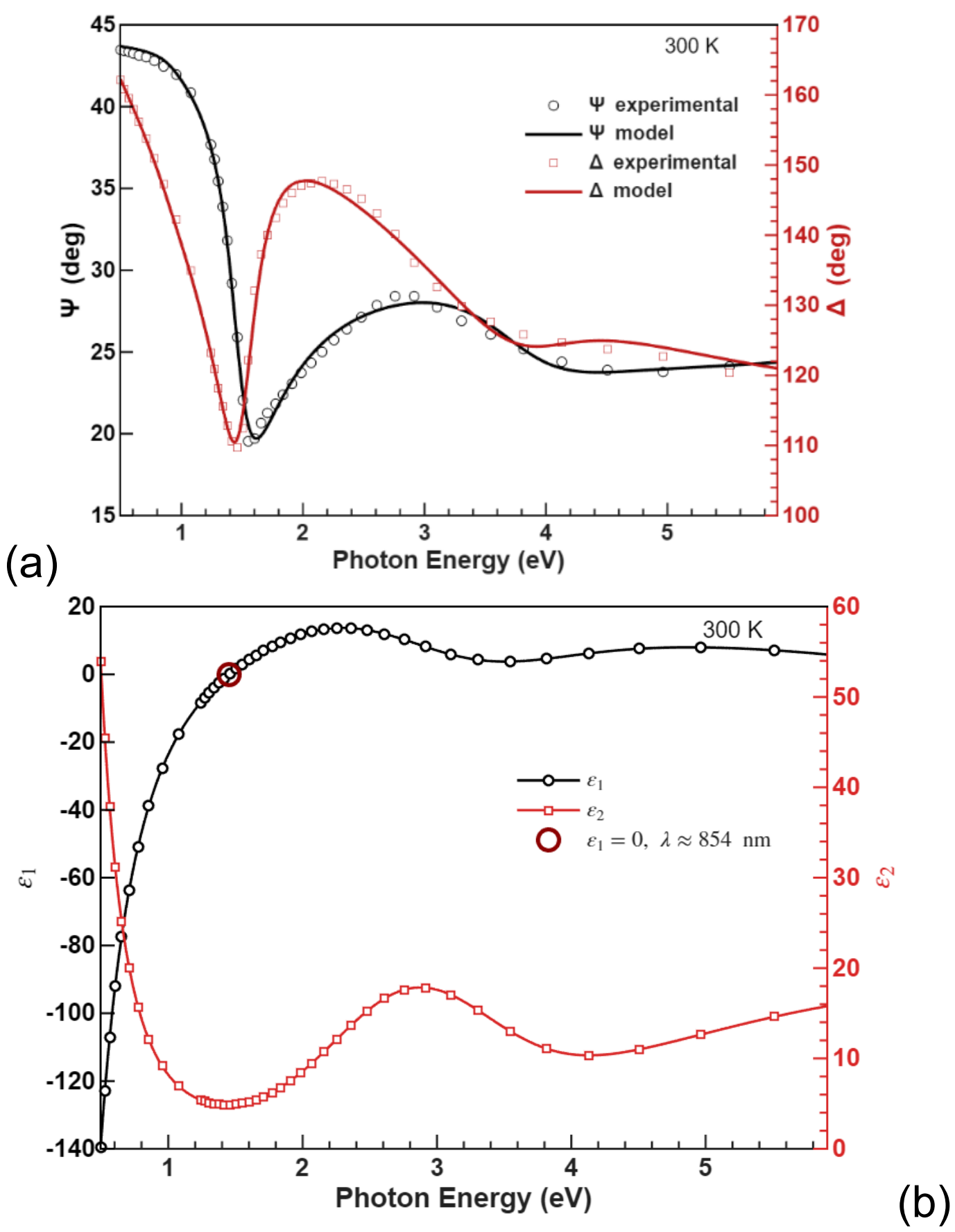

Figure SI-A: At room temperature (~ 300 K) (a) spectroscopic ellipsometry data for (100) $ScB_2$; Experimental $\psi$ and $\Delta$ spectra are shown together with the corresponding fits obtained using a Drude plus two Gaussian oscillator optical model. (b) real ($\varepsilon_1$) and imaginary parts ($\varepsilon_2$) of dielectric constants.

**B. Rocking curves from high resolution XRD**

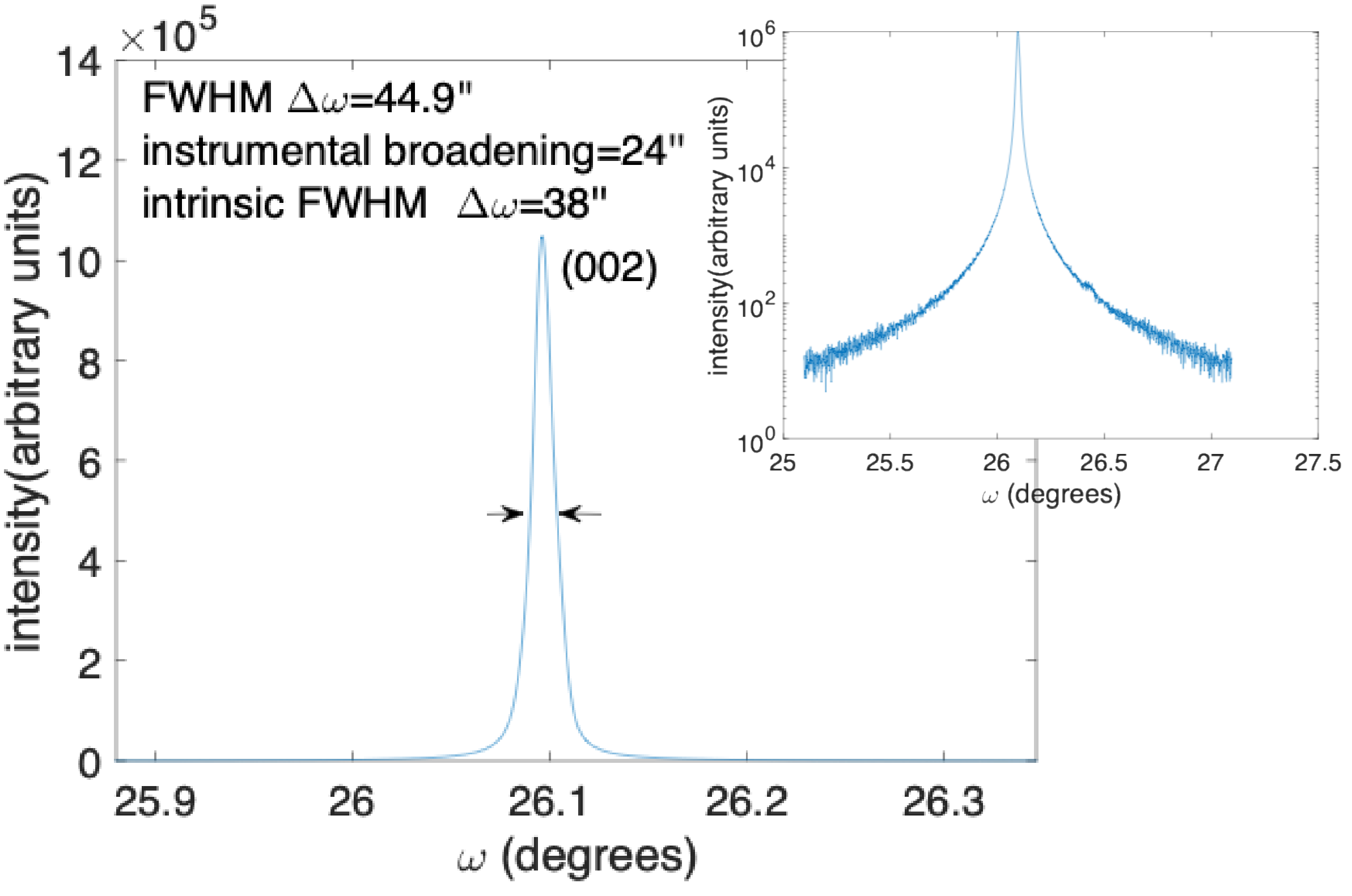

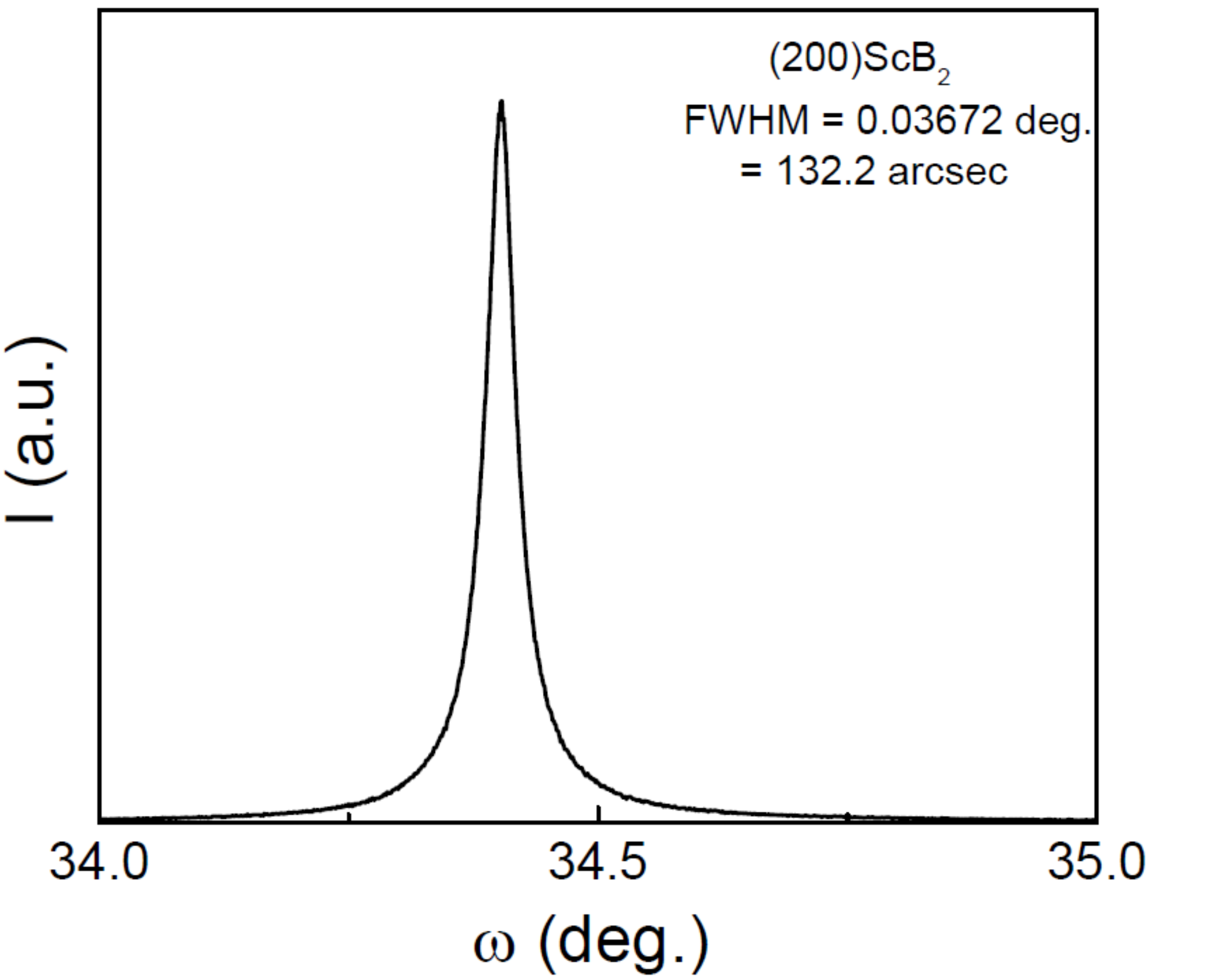

**C. Band structure calculations**

The electronic density of states (DOS) of $ScB_2$ was calculated with the SCAN functional at the relaxed equilibrium lattice parameters using a 12×12×10 k-mesh for Brillouin zone sampling.

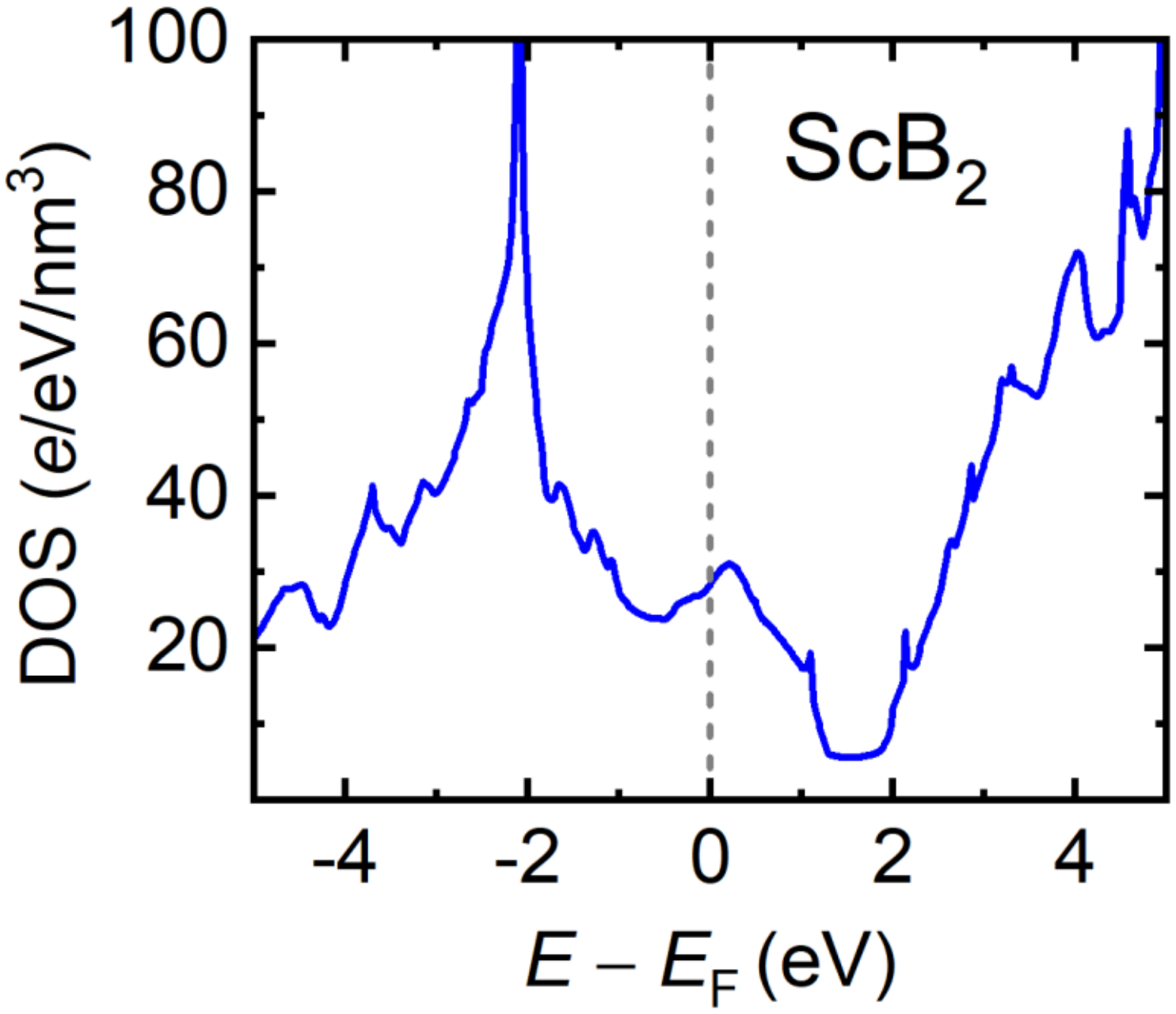

## D. Atomic force microscopy (AFM)

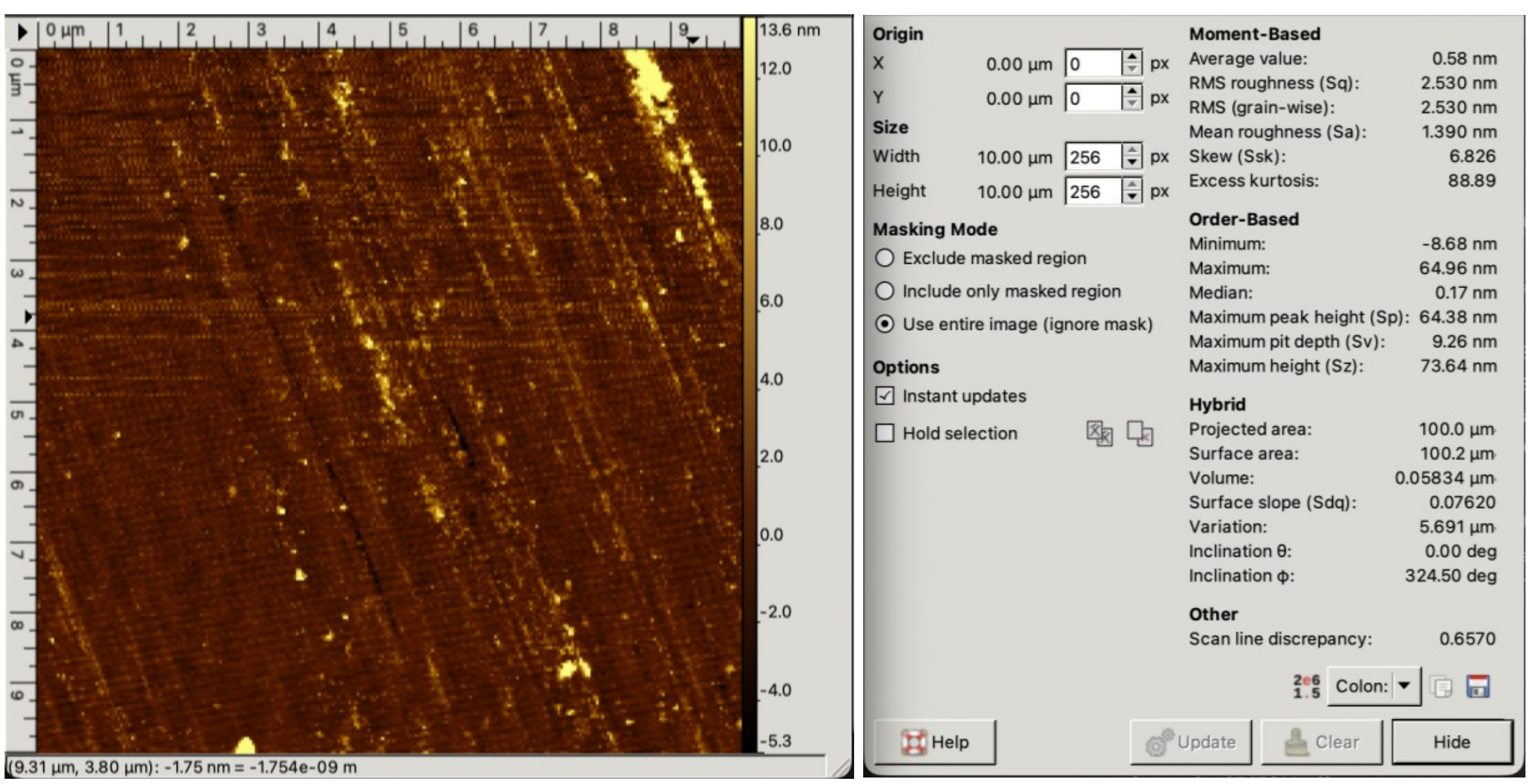

Figure D (a) Atomic force micrograph of optically polished $ScB_2$ showing 2.5nm rms roughness for $10\mu m \times 10\mu m$ area.

## E. Scanning Electron Microscopy (SEM) and Electron Backscatter Diffraction (EBSD)

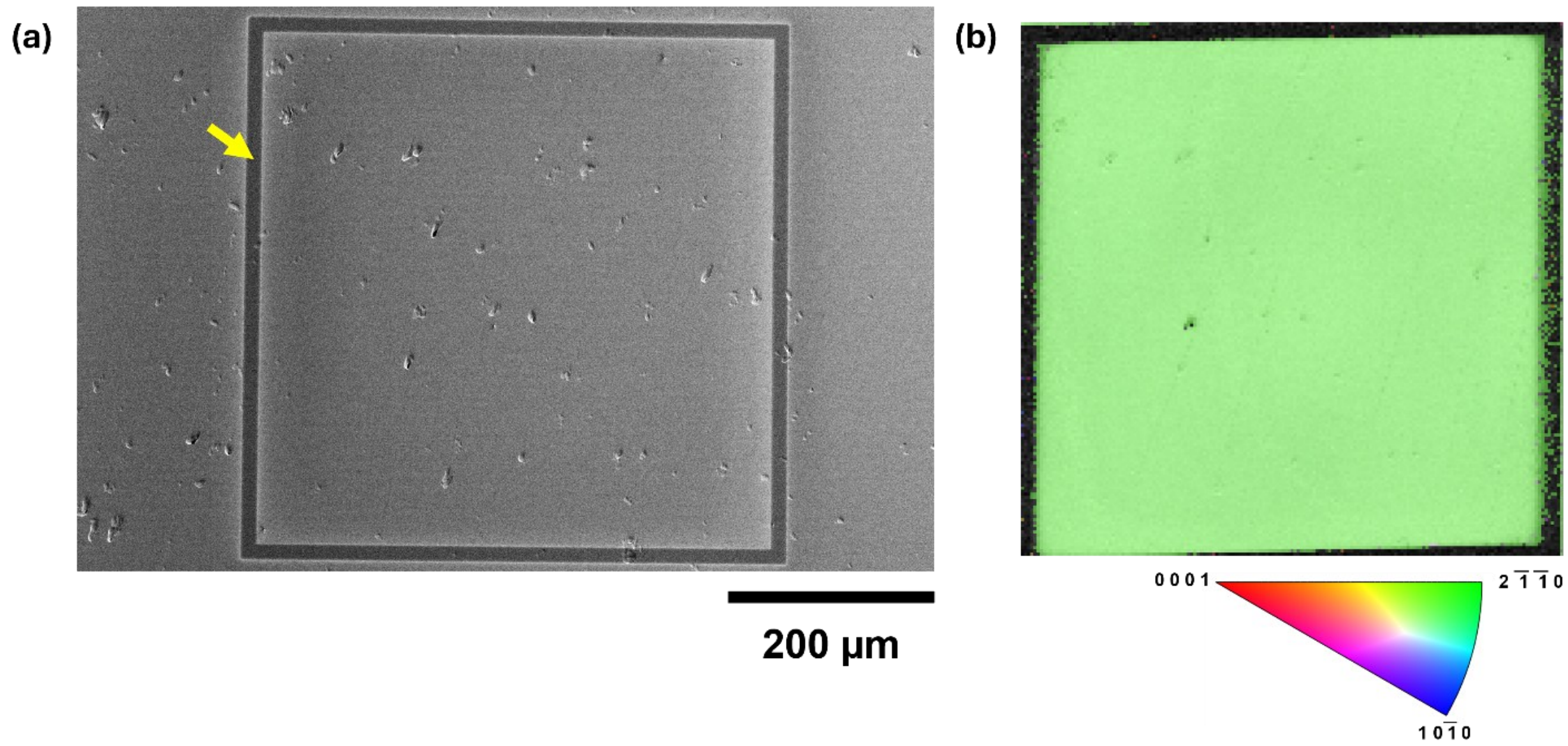

Figure E (a) Secondary Electron SEM image at 2 kV 0.2 nA of the $ScB_2$ surface after Argon BIB polishing and PFIB milling of a rectangular fiducidal (marked with yellow arrow). (b) EBSD Inverse Pole Figure (IPF, color) + Image Quality (IQ, grayscale) map relative to the normal direction of PFIB fiducial marked region shown in (a) of $ScB_2$ cut along the growth direction at a 3 µm step size. Uniform color in the IPF map is consistent with a single crystal orientation of [-12-11] and brightness of the IQ in the region inside fiducial is consistent with a smooth surface and high pattern quality.

## F. Small crystals/polycrystalline samples grown under Sc-rich conditions

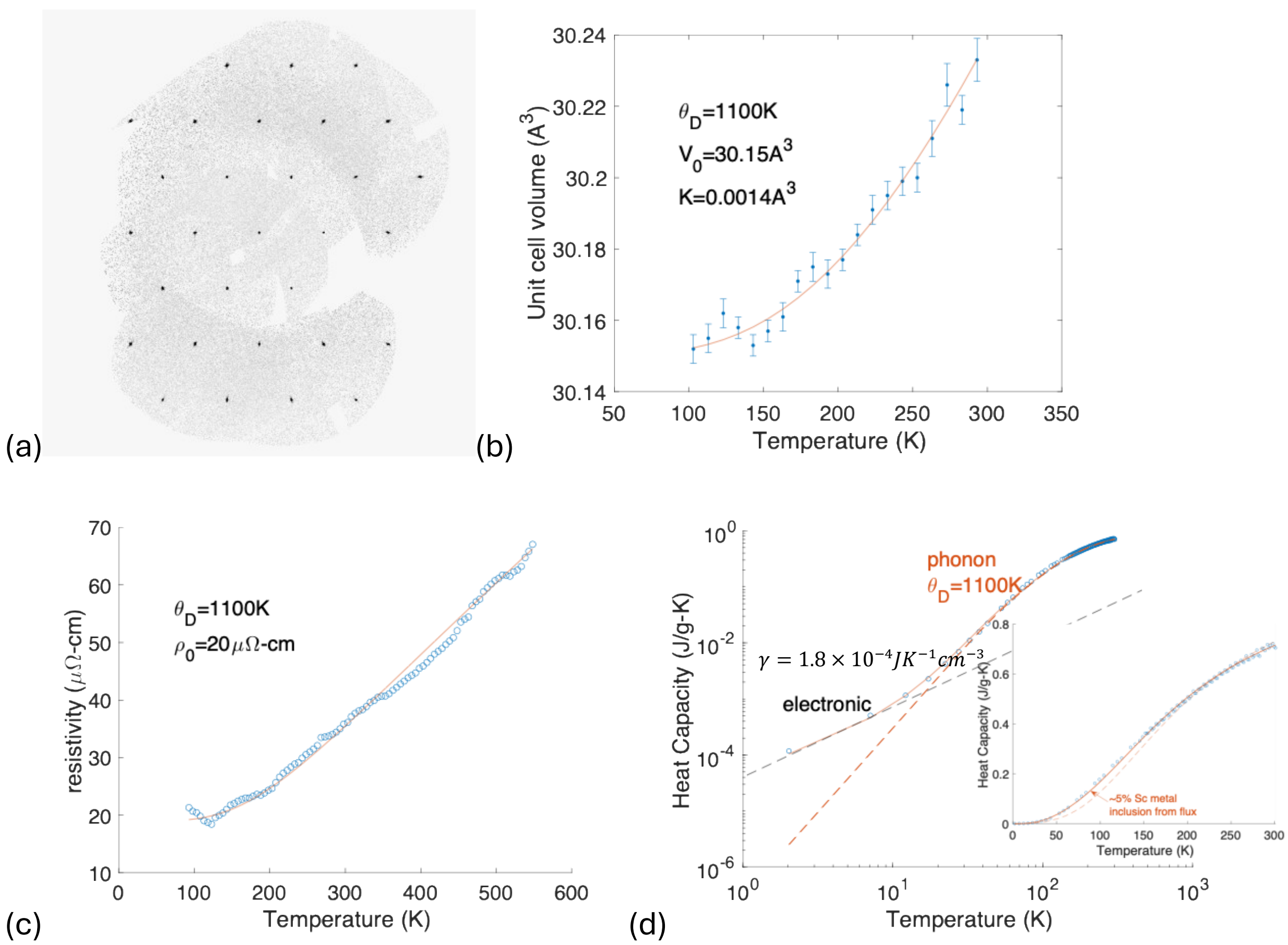

Figure SI-F (a) Room temperature single crystal x-ray diffraction (SCXRD) pattern for (hk1) reflections from which the crystal structure of $ScB_2$ was refined (b) Temperature dependence of unit cell volume indicating a Debye temperature $\Theta_D$ ~1100K (c) Temperature dependence of bulk resistivity showing excellent agreement with Bloch-Gruneisen behavior with a $\theta_D \sim 1100K$ in agreement with lattice expansion in (b) (d)Temperature dependence of heat capacity showing a Debye temperature $\Theta_D$~1100K in agreement with lattice expansion in (b), and resistivity in (d).

**SCXRD Details:** The crystal was preliminarily glued on a broken nylon loop to prevent any undesirable crystal movement at higher temperature (especially above 270 K). The data collection temperature was regulated by an Oxford Cryostream 1000. The data collection strategy (29-ω scans, 1815 frames) was employed for all measurements. The crystal was subsequently cooled and heated in ±10 K increments at a rate of ±2 K/min over the 103–293 K range. Before each measurement, a 5-minute stabilization period was allowed to ensure thermal equilibrium. In the 1$^{st}$ cycle, 20 data points were collected while cooling from 293 K to 103 K. In the second cycle, while heating from 103 to 293 K, 19 data points were collected respectively.

## G. Computational Methods for Electronic and Phononic Contributions to Thermal Conductivity in $ScB_2$

**<u>First principles electronic transport</u>:** First-principles calculations were carried out using density-functional theory (DFT) and density-functional perturbation theory (DFPT) methods, as implemented in the Quantum Espresso suite. Optimized norm-conserving Vanderbilt pseudopotentials with Perdew–Burke–Ernzerhof exchange-correlation functional was used to execute these calculations [G1-4]. Subsequently, we also utilized Electron–Phonon Wannier (EPW) code [G5,6], a transport module where the temperature-dependent electronic properties were computed through the interpolation of electronic bands, phonons and electron–phonon matrix on a densely sampled Brillouin-Zone mesh with the help of maximally localized Wannier functions, as implemented in the EPW code [G5,7,8].

The computations done using EPW code required us to first calculate the quantities such as the electronic states, phonon dispersions, and electron–phonon matrix elements on a coarse electronic and phonon wave-vector grids, which in our case, were chosen to be $12 \times 12 \times 12$ and $6 \times 6 \times 6$ respectively. These quantities calculated on a coarse grid were then interpolated on to a denser $72 \times 72 \times 72$ $\mathbf{k}$ and $36 \times 36 \times 36$ $\mathbf{q}$ grids, respectively, by utilizing the maximally localized Wannier functions.

Once the electronic bands, phonon dispersions, and the electron–phonon matrix elements were Wannier interpolated, and electronic thermal conductivity was calculated by iteratively solving the linearized *ab initio* Boltzmann transport equation [G6]. A major difference between the self-energy relaxation-time approximation (SERTA) and the iterative solution is the that the iterative solution incorporates the electron-phonon scattering induced carrier redistribution, thereby including the scattering-in contribution to the Boltzmann transport equation. Therefore, the electronic distribution's first order response to an applied electric field applied along the Cartesian direction $\beta$ under the above-mentioned convention is written as

$$\begin{aligned}\frac{\partial f_{n\mathbf{k}}}{\partial E_\beta} = &\quad e v_{n\mathbf{k},\beta} \frac{\partial f^0_{n\mathbf{k}}}{\partial \varepsilon_{n\mathbf{k}}} \tau_{n\mathbf{k}} \\ &+ \frac{2\pi\tau_{n\mathbf{k}}}{\hbar} \sum_{m\nu} \int \frac{d\mathbf{q}}{\Omega_{\mathrm{BZ}}} |g_{mn\nu}(\mathbf{k},\mathbf{q})|^2 \frac{\partial f_{m,\mathbf{k+q}}}{\partial E_\beta} \mathcal{P}_{mn\nu}(\mathbf{k},\mathbf{q}),\end{aligned}$$

where the terms related to the phonon absorption and emission are combined into

$$\begin{aligned}\mathcal{P}_{mn\nu}(\mathbf{k},\mathbf{q}) = &\quad \left(n_{\mathbf{q}\nu} + 1 - f^0_{n\mathbf{k}}\right)\delta\left(\varepsilon_{n\mathbf{k}} - \varepsilon_{m,\mathbf{k+q}} + \hbar\omega_{\mathbf{q}\nu}\right) \\ &+ \left(n_{\mathbf{q}\nu} + f^0_{n\mathbf{k}}\right)\delta\left(\varepsilon_{n\mathbf{k}} - \varepsilon_{m,\mathbf{k+q}} - \hbar\omega_{\mathbf{q}\nu}\right).\end{aligned}$$

Here, $f_{n\mathbf{k}}$ and $f_{n\mathbf{k}}^0$ denote the nonequilibrium and equilibrium electronic distribution functions, respectively; $v_{n\mathbf{k},\beta}$ is the electronic band velocity; $\tau_{n\mathbf{k}}$ is the state-dependent electron–phonon relaxation time; $g_{mn\nu}(\mathbf{k},\mathbf{q})$ is the electron–phonon matrix element coupling the initial electronic state $(n,\mathbf{k})$ to the final state $(m,\mathbf{k}+\mathbf{q})$ through a phonon of branch $\nu$ and wavevector $\mathbf{q}$; and $n_{\mathbf{q}\nu}$ is the Bose–Einstein occupation of that phonon mode.

The state-dependent electron–phonon scattering rate entering the transport calculation is obtained from the corresponding phonon absorption and emission processes. In compact form,

$$\tau_{n\mathbf{k}}^{-1} = \frac{2\pi}{\hbar}\sum_{m\nu}\int \frac{d\mathbf{q}}{\Omega_{\mathrm{BZ}}}|g_{mn\nu}(\mathbf{k},\mathbf{q})|^2 \mathcal{Q}_{mn\nu}(\mathbf{k},\mathbf{q}),$$

where

$$\begin{aligned}\mathcal{Q}_{mn\nu}(\mathbf{k},\mathbf{q}) = \quad & \left(n_{\mathbf{q}\nu} + 1 - f_{m,\mathbf{k}+\mathbf{q}}^0\right)\delta\left(\varepsilon_{n\mathbf{k}} - \varepsilon_{m,\mathbf{k}+\mathbf{q}} - \hbar\omega_{\mathbf{q}\nu}\right) \\ & +\left(n_{\mathbf{q}\nu} + f_{m,\mathbf{k}+\mathbf{q}}^0\right)\delta\left(\varepsilon_{n\mathbf{k}} - \varepsilon_{m,\mathbf{k}+\mathbf{q}} + \hbar\omega_{\mathbf{q}\nu}\right).\end{aligned}$$

The electrical conductivity tensor can be evaluated from the field dependent electronic distribution, after the convergence of the iterative BTE [G6]

$$\sigma_{\alpha\beta} = -\frac{e}{\Omega}\sum_{n}\int \frac{d\mathbf{k}}{\Omega_{\mathrm{BZ}}} v_{n\mathbf{k},\alpha}\frac{\partial f_{n\mathbf{k}}}{\partial E_\beta},$$

where $\Omega$ is the primitive-cell volume and $\Omega_{\mathrm{BZ}}$ is the Brillouin-zone volume.

The state-resolved quantities are then used to the reconstruct the electrical conductivity tensor through

$$\sigma_{\alpha\beta} = \frac{e^2}{V}\sum_{n\mathbf{k}} w_{n\mathbf{k}}\ v_{n\mathbf{k},\alpha}v_{n\mathbf{k},\beta}\tau_{n\mathbf{k}}\left(-\frac{\partial f}{\partial\varepsilon}\right),$$

where $V$ is the primitive-cell volume and $w_{n\mathbf{k}}$ denotes the Brillouin-zone integration weight. The estimation of the first energy moment with respect to the thermoelectric transport tensor was done by

$$(\sigma S)_{\alpha\beta} = \frac{e}{VT}\sum_{n\mathbf{k}} w_{n\mathbf{k}}\ \xi_{n\mathbf{k}}\, v_{n\mathbf{k},\alpha}v_{n\mathbf{k},\beta}\tau_{n\mathbf{k}}\left(-\frac{\partial f}{\partial\varepsilon}\right),$$

and the second energy moment, or electronic heat-transport tensor, as

$$K_{\alpha\beta} = \frac{1}{VT}\sum_{n\mathbf{k}} w_{n\mathbf{k}}\ \xi_{n\mathbf{k}}^2\, v_{n\mathbf{k},\alpha} v_{n\mathbf{k},\beta} \tau_{n\mathbf{k}} \left(-\frac{\partial f}{\partial \varepsilon}\right).$$

In the constraint of open-circuit conditions, the electronic thermal conductivity is calculated from

$$\boldsymbol{\kappa}_e = \mathbf{K} - T(\boldsymbol{\sigma}\mathbf{S})\boldsymbol{\sigma}^{-1}(\boldsymbol{\sigma}\mathbf{S}),$$

where the second term pertains to the thermoelectric contribution. In the case of zero-electrical-current condition, this thermoelectric contribution must be discarded. By inverting the electrical conductivity tensor, we can determine the electrical resistivity tensor,

$$\boldsymbol{\rho} = \boldsymbol{\sigma}^{-1}.$$

The calculation of Fermi-window-weighted carrier velocities and effective lifetimes help shed some light on the minuscule onset of trends in electronic transport. The root-mean-square velocity along direction $\alpha$ was defined as

$$v_{\mathrm{RMS},\alpha} = \left[\frac{\sum_{n\mathbf{k}} w_{n\mathbf{k}}\ v_{n\mathbf{k},\alpha}^2 \left(-\frac{\partial f}{\partial \varepsilon}\right)}{\sum_{n\mathbf{k}} w_{n\mathbf{k}} \left(-\frac{\partial f}{\partial \varepsilon}\right)}\right]^{1/2},$$

whereas the transport-weighted effective lifetime was evaluated as

$$\tau_{\mathrm{eff},\alpha} = \frac{\sum_{n\mathbf{k}} w_{n\mathbf{k}}\ v_{n\mathbf{k},\alpha}^2 \tau_{n\mathbf{k}} \left(-\frac{\partial f}{\partial \varepsilon}\right)}{\sum_{n\mathbf{k}} w_{n\mathbf{k}}\ v_{n\mathbf{k},\alpha}^2 \left(-\frac{\partial f}{\partial \varepsilon}\right)}.$$

The evaluation of these parameters was done for the sole purpose of utilizing them as microscopic indicators of electronic states that are involved in heat and charge transport and not make any alterations to the scattering rates that were calculated through the EPW code.

**First-principles lattice thermal transport:** To obtain the thermal conductivity values, the linearized phonon Boltzmann transport equation was solved iteratively using ShengBTE [G9-10]. To execute these calculations, we require harmonic (second-order) and anharmonic (third-order) interatomic force constants (IFCs), which are evaluated through first-principles calculations. The relaxation-time approximation form of the lattice thermal conductivity tensor incorporating the different phonon modes is expressed as

$$\kappa_{\mathrm{L}}^{\alpha\beta} = \frac{1}{N_{\mathbf{q}} V_{\mathrm{c}}} \sum_{\mu} C_\mu\, v_\mu^\alpha v_\mu^\beta \tau_\mu,$$

where the composite index $\mu = (\mathbf{q}, \nu)$ represents a phonon mode whose wave vector is $\mathbf{q}$ and its branch index $\nu$. The number of wave vectors utilized for the sampling of the Brillouin Zone is denoted by $N_{\mathbf{q}}$ and $V_{\mathrm{c}}$ corresponds to the volume of the primitive-cell.

Other parameters within the lattice thermal conductivity equation includes the mode-resolved heat capacity denoted by $C_\mu$, the $\alpha$ component of the group velocity of the phonon represented by $v_\mu^\alpha$ and the phonon lifetime which is indicated by $\tau_\mu$. Specifically, all the calculations done in this computational work was done by iteratively solving the BTE, rather than relying on the single-mode relaxation-time approximation.

For the convergence of the ground state electronic charge densities using the first-principles calculations, a plane-wave cutoff energy of 50 Ry was applied to a $11 \times 11 \times 10$ Monkhorst–Pack $\mathbf{k}$-point grid. The computation of the harmonic dynamical matrices were done on a $11 \times 11 \times 10$ uniform phonon wave-vector grid. By Fourier transforming the dynamical matrices, we acquire the second-order interatomic force constants (IFCs), which is then utilized to visualize phonon dispersion and produce quantities such as eigenvectors and group velocities.

The construction of third-order IFCs was achieved by implementing the finite-displacement real-space supercell approach using the thirdorder.py code, which comes with the distribution of ShengBTE [G10], where a supercell size of $4 \times 4 \times 4$ containing 192 atoms was used. The consideration of the third-order interactions was limited to the tenth-nearest-neighboring shell.

A Brillouin-zone with a dense $\mathbf{q}$-point grid of $22 \times 22 \times 20$ was employed to compute the converged lattice thermal-conductivity tensor for $ScB_2$.

To examine the effect of electron–phonon coupling on the lattice thermal conductivity, we calculate the mode-resolved electron–phonon scattering rates using the EPW code. Matthiesen’s rule is then applied to combine these phonon-phonon and electron-phonon scattering rates

$$\frac{1}{\tau_\mu^{\mathrm{3ph+ep}}} = \frac{1}{\tau_\mu^{\mathrm{3ph}}} + \frac{1}{\tau_\mu^{\mathrm{ep}}},$$

where $\tau_\mu^{\mathrm{3ph}}$ and $\tau_\mu^{\mathrm{ep}}$ are the mode-resolved lifetimes associated with intrinsic three-phonon and phonon–electron scattering, respectively. Another equivalent way to compute the combined scattering rates is given by

$$\Gamma_\mu^{\mathrm{3ph+ep}} = \Gamma_\mu^{\mathrm{3ph}} + \Gamma_\mu^{\mathrm{ep}},$$

Isolating the effect of electron–phonon coupling is achieved by solving the phonon BTE under the two scattering conditions. In the first scenario, when we include the intrinsic three-phonon scattering, we obtain $\kappa_{\mathrm{L}}^{\mathrm{3ph}}$, whereas in the second scenario, if we combine the phonon–phonon and electron–phonon scattering rates, then we obtain $\kappa_{\mathrm{L}}^{\mathrm{3ph+ep}}$. Therefore, the reduction in the lattice thermal conductivity due to electron–phonon coupling is isolated by taking the difference of the two thermal conductivity tensors. Here $\kappa_{\mathrm{L}}^{\mathrm{3ph}}$ is the intrinsic limit of the thermal conductivity tensor in a three-phonon scattering process, wereas $\kappa_{\mathrm{L}}^{\mathrm{3ph+ep}}$ is the intrinsic limit of the thermal conductivity tensor that accounts for both the three-phonon as well as the electron–phonon scattering processes.

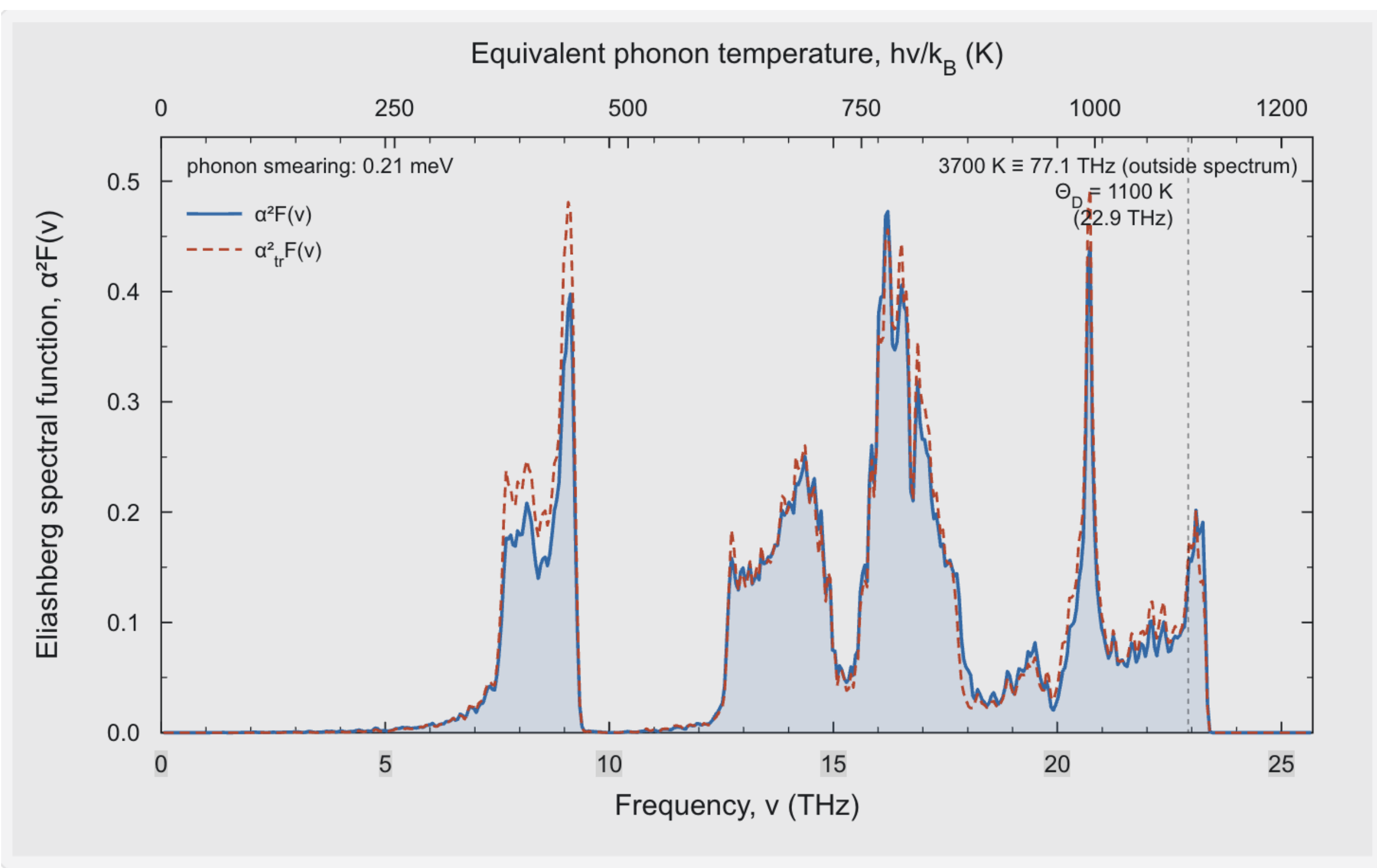

**Figure G: Eliashberg spectral function for electron-phonon coupling from first principles.**